\documentclass[a4paper]{jpconf}

\usepackage{amsmath}
\usepackage{amssymb}

\usepackage{geometry}
\usepackage{graphicx}
\usepackage{citesort}

\begin{document}

\title{Semi-analytical model of acoustic-wave generation by a laser line
pulse in an optically absorptive isotropic cylinder}
\author{D S\'egur$^1$, A L Shuvalov$^1$, Y Pan$^2$, B Audoin$^1$}
\address{$^1$ Laboratoire de M\'ecanique Physique, UMR CNRS 8469,
  Universit\'e Bordeaux 1, 33405 Talence, France}
\address{$^2$ Institute of Acoustics, Tongji University, Shanghai, China}
\ead{d.segur@lmp.u-bordeaux1.fr}

\begin{abstract}
Semi-analytical model for calculating acoustic response to a line-focused
laser pulse in an optically absorptive isotropic cylinder is proposed
and implemented. It is assumed that the laser input is absorbed over
the volume and thus creates a radially distributed thermoelastic source. Closed-form solution is
obtained in the Fourier domain. Two inverse Fourier transforms in frequency
and circumferential wave number yield the sought waveforms of acoustic
response in the time-space domain. Numerical simulation is compared to the
calculation based on a surface dipole source. The two signals have
essentially different shapes of the wave-arrival peaks when the optical penetration is large enough.

\medskip

\end{abstract}

\section{Introduction}

The laser ultrasonics technique is a unique tool for acoustic-wave
generation in objects with a curved surface. Much work has been done by this
means for generating surface and bulk acoustic waves in opaque cylindrical
targets with a view of their non-destructive testing and evaluation, see
recent publications \cite{Clorennec:2002,Clorennec:2003,Pan:2003,Pan:2004}
and the bibliography therein. There exists a well developed theory for the
case when, at a given laser wavelength, the target (either cylindrical or
planar) is opaque and hence the source is located at the surface. Given so,
the ablation model applies for high enough local energy such that causes
ablation at the laser-heated region of the surface, and the dipole
(Scruby's) model applies for relatively low incoming energy \cite%
{Scruby:1990}. In this latter, thermoelastic regime, the dilatation of an
infinitely small volume adjacent to the surface produces a surface source
equivalent to a force dipole. The dipole strength has been derived in \cite%
{Royer:2001,Arias:2003}.

Thermoelastic regime for the optically absorptive materials implies that
the laser input penetrates into a target and creates a volume distribution
of sources therein. Hence the assumption of a surface source is no longer
appropriate. Laser generation of acoustic waves in optically absorptive materials has
been considered for the case of a planar target (plate) \cite{Dubois:1994}. The next
natural step is extending laser ultrasonics treatment to optically absorptive
cylindrical targets. To the best of the authors' knowledge, such a case has
not yet been studied. The present paper proposes a two-dimensional (2D)
semi-analytical model for calculating acoustic response to a line-focused
laser pulse in an optically absorptive isotropic cylinder. Thermoelastic regime of
generation is assumed and a typical case of materials with negligible
thermal conductivity is treated. Closed-form solution is obtained in the
Fourier domain to the inhomogeneous ordinary differential system with a
radially distributed thermoelastic source and with the traction-free
boundary condition at the cylinder surface. Performing two inverse Fourier
transforms yields the sought waveforms of acoustic response in the
time-space domain. This calculation is carried out in the paper for two
different source settings. One of them formally treats the case of an opaque
cylinder by assuming a vanishingly small optical penetration or, which is
equivalent, a buried source tending to the cylinder surface. The other
considers the actual case of an optically absorptive cylinder, where the exponentially
decaying law of optical absorption produces a distributed source with large
enough radial penetration length. The results obtained for these two
settings are compared with the calculation based on Scruby's model of a
surface dipole source. It is observed that the waveforms, computed for a
buried source tending to the surface, agree very well with those following
for Scruby's surface-source model, whereas the waveforms, computed for a
distributed source, have essentially different shapes comparatively to the
prediction following for the surface source. The former observation serves
as a test for our model, while the latter highlights the effect of
laser-energy penetration into the bulk of an optically absorptive cylinder.

\section{Background}

Consider an infinite homogenous isotropic elastic cylinder of radius $R$,
whose surface is heated by the line-focused laser pulse independent of the
axial coordinate $z$. The laser energy is supposed to be low enough to
preclude ablation. It is assumed that viscoelasticity, thermal conduction and the
contribution of deformation to the entropy change may be disregarded. The
thermoelastic generation of acoustic wave is therefore described by the
following simplified system of governing equations:%
\begin{equation}
\mathbf{\nabla \cdot \sigma }=\rho \mathbf{\ddot{u}};\ \mathbf{\sigma }%
=\lambda \mathrm{tr}\left( \mathbf{\varepsilon }\right) \mathbf{I}+2\,\mu 
\mathbf{\varepsilon -}\beta T\mathbf{I;\ }\rho c_{p}\dot{T}=q\,,
\label{eq::coupled_thermoelastic_equations}
\end{equation}%
where $\mathbf{\sigma }$ is the stress tensor including mechanical and
thermal stresses, $\mathbf{\varepsilon }$ is the strain tensor, $\mathbf{u}$
is the displacement, $\rho $ is the density, $\lambda $ and $\mu $ are the
Lam\'{e} coefficients, $\beta =(3\lambda +2\mu )\,\alpha _{T}$ is the
thermal modulus ($\alpha _{T}$ is the dilatation coefficient), $T$ is the
temperature rise, $c_{p}$ is the specific heat, $q$ is the power density of
the line laser source, $\mathbf{I}$ is the identity matrix and $\mathrm{tr}$
denotes the trace.

The heat input $q(r,\theta ,t)$ creates a volume distribution of thermal
sources inside the cylinder through the optical absorption. Assuming normal
incidence of the Gaussian laser beam \cite{Hecht:2002} and the exponential
fall off of the incoming energy $E$, it follows that 
\begin{equation}
q(r,\theta ,t)=\,\alpha \,E\,g(\theta )\,\delta (t)\,e^{-\alpha \,(R-r)},
\label{3}
\end{equation}%
where 
\begin{equation}
g(\theta )=\frac{1}{\gamma }\sqrt{\frac{4\ln 2}{\pi }}e^{-\left( 4\,\ln
2\,\right) \frac{\theta ^{2}}{\gamma ^{2}}},\ \gamma =2\,\arctan \frac{b}{2R}
\end{equation}%
describes the angular extent of the beam with width $b$. By (\ref%
{eq::coupled_thermoelastic_equations})$_{3}$ and (\ref{3}), 
\begin{equation}
T(r,\theta ,t)=\frac{\alpha E}{\rho c_{p}}\,g(\theta )\,H(t)\,e^{-\alpha
(R-r)},  \label{eq::temperature_field}
\end{equation}%
where $H$ is the Heaviside function. The temperature field (\ref%
{eq::temperature_field}) specifies the thermal stress $\sigma ^{\left(
T\right) }=-\beta T$ in (\ref{eq::coupled_thermoelastic_equations}).

Thus, the thermoelastic problem is reduced to solving the 2D plane-strain
equation of motion for the displacement $\mathbf{u}(r,\theta ,t)=u_{r}%
\mathbf{e}_{r}+u_{\theta }\mathbf{e}_{\theta }$ in the cylinder
cross-section $\left( \mathbf{e}_{r},\mathbf{e}_{\theta }\right) $.
Divergence of the thermal stress plays the role of a source term. The
equation is complemented by the traction-free boundary condition (BC) $%
\mathbf{e}_{r}\mathbf{\sigma }=\mathbf{0}$ at the cylinder surface $r=R$.

\section{Solution in the Fourier domain}

\subsection{Setup}

Consider the above 2D problem in the Fourier domain $\left\{ r,\nu ,\omega
\right\} $, where $\nu $ is the circumferential wave number and $\omega $ the
angular frequency. Thereafter in this section the factor $e^{j\left( \nu
\theta -\omega t\right) }$ is omitted, the same notations for the
transformed functions are used, and their dependence on $\nu ,\omega $ is
understood and suppressed. According to (\ref%
{eq::coupled_thermoelastic_equations}) and (\ref{eq::temperature_field}), we
write%
\begin{equation}
\begin{array}{c}
\mathbf{\sigma }\left( r\right) =\mathbf{\sigma }^{\left( m\right) }\left(
r\right) +\sigma ^{\left( T\right) }\left( r\right) \mathbf{I,} \\ 
\mathbf{\sigma }^{\left( m\right) }\left( r\right) =\lambda \mathrm{tr}\left[
\mathbf{\varepsilon }\left( r\right) \right] \mathbf{I}+2\,\mu \mathbf{%
\varepsilon }\left( r\right) ,\ \sigma ^{\left( T\right) }\left( r\right)
=-\beta T_{0}e^{-\alpha (R-r)},%
\end{array}
\label{P}
\end{equation}%
where $T_{0}\left( \sim \alpha \right) $ involves double transform of (\ref%
{eq::temperature_field}) in $\theta $ and $t.$

The equation of motion in the Fourier domain reduces to the ordinary
differential system in $\mathbf{u}\left( r\right) $ with the source term
provided by $\mathbf{\nabla }\sigma ^{\left( T\right) }\left( r\right) .$
Using Helmholtz decomposition, this equation may be re-cast in the form of
two uncoupled Bessel equations for the longitudinal and shear elastic
potentials $\varphi \left( r\right) $ and $\psi \left( r\right) $, 
\begin{equation}
\left( 
\begin{array}{cc}
\mathbf{\nabla }^{2}+k_{L}^{2} & 0 \\ 
0 & \mathbf{\nabla }^{2}+k_{T}^{2}%
\end{array}%
\right) \left( 
\begin{array}{c}
\varphi \left( r\right) \\ 
\psi \left( r\right)%
\end{array}%
\right) =\left( 
\begin{array}{c}
be^{-\alpha (R-r)} \\ 
0%
\end{array}%
\right) ,  \label{P3}
\end{equation}%
where $k_{L}^{2}=\rho \omega ^{2}/\left( \lambda +2\mu \right) ,\
k_{T}^{2}=\rho \omega ^{2}/\mu $ and $b=\beta T_{0}/\left( \lambda +2\mu
\right) .$ The traction-free BC, which couples $\varphi \left( r\right) $
and $\psi \left( r\right) ,$ is 
\begin{equation}
\left( 
\begin{array}{c}
\sigma _{rr}^{\left( m\right) }\left( R\right) \\ 
\sigma _{r\theta }^{\left( m\right) }\left( R\right)%
\end{array}%
\right) +\left( 
\begin{array}{c}
-\beta T_{0} \\ 
0%
\end{array}%
\right) =\mathbf{0,}  \label{P5}
\end{equation}%
where 
\begin{equation}
\left( 
\begin{array}{c}
\sigma _{rr}^{\left( m\right) }\left( R\right) \\ 
\sigma _{r\theta }^{\left( m\right) }\left( R\right)%
\end{array}%
\right) =\mathbf{S}\left( R\right) \left( 
\begin{array}{c}
\varphi \left( R\right) \\ 
\psi \left( R\right)%
\end{array}%
\right) ,\ \mathbf{S}\left( r\right) =2\mu \left( 
\begin{array}{cc}
\frac{\lambda }{2\mu }\mathbf{\nabla }^{2}+\,\frac{\mathrm{d}^{2}}{\mathrm{d}%
r^{2}}\, & \frac{\jmath \nu }{r}(\frac{\mathrm{d}}{\mathrm{d}r}-\frac{1}{r}) \\ 
\,\frac{\jmath \nu }{r}(\frac{\mathrm{d}}{\mathrm{d}r}-\frac{1}{r})\, & \frac{1}{2}%
\mathbf{\nabla }^{2}\,-\frac{\mathrm{d}^{2}}{\mathrm{d}r^{2}}%
\end{array}%
\right) .  \label{P4}
\end{equation}

For the future use, introduce the following abridged notations of the values
taken at the boundary $r=R$: 
\begin{equation}
K_{L,T}\equiv k_{L,T}R,\ B_{T,L}\equiv \frac{K_{T,L}J_{\nu }^{\prime }\left(
K_{T,L}\right) }{J_{\nu }\left( K_{T,L}\right) },\ ^{\prime }\equiv \left[ 
\frac{\mathrm{d}J_{\nu }\left( x\right) }{\mathrm{d}x}\right] _{x=K_{L,T}}.
\label{P11.1}
\end{equation}%
Note that expressing the results in terms of $B_{T,L}$ is helpful for
numerical implementation, because the exponential growth of the Bessel
functions of the order $\nu \gg K_{L,T}\,\ $is avoided this way (see \cite%
{Gespa:1987}).

\subsection{Green function}

First, we solve an auxiliary problem for a buried point source at $r_{0}$ ($%
0\leq r_{0}<R$) under the homogeneous BC on the mechanical stresses (see (%
\ref{P4})$_{1}$),%
\begin{equation}
\left( 
\begin{array}{cc}
\mathbf{\nabla }^{2}+k_{L}^{2} & 0 \\ 
0 & \mathbf{\nabla }^{2}+k_{T}^{2}%
\end{array}%
\right) \left( 
\begin{array}{c}
\Phi \left( r|r_{0}\right)  \\ 
\Psi \left( r|r_{0}\right) 
\end{array}%
\right) =\left( 
\begin{array}{c}
bR^{2}\frac{\delta \left( r-r_{0}\right) }{r} \\ 
0%
\end{array}%
\right) ,\ \ \mathbf{S}\left( R\right) \left( 
\begin{array}{c}
\Phi \left( R|r_{0}\right)  \\ 
\Psi \left( R|r_{0}\right) 
\end{array}%
\right) _{r_{0}<R}=\mathbf{0,}  \label{P7}
\end{equation}%
where the factor $R^{2}$ in (\ref{P7})$_{1}$ is added to merely keep
physical dimensions intact. The partial solution to the given inhomogeneous
Bessel equation on $\Phi \left( r|r_{0}\right) $ is \cite{Kamke:1971}%
\begin{equation}
g\left( r|r_{0}\right) =bR^{2}\frac{\pi }{2}\left[ J_{\nu }\left(
k_{L}r\right) Y_{\nu }\left( k_{L}r_{0}\right) H\left( r_{0}-r\right)
+Y_{\nu }\left( k_{L}r\right) J_{\nu }\left( k_{L}r_{0}\right) H\left(
r-r_{0}\right) \right] ,  \label{P9}
\end{equation}%
where $J_{\nu }$ and $Y_{\nu }$ are the Bessel functions of the first and
second kind. Thus the solution to (\ref{P7}) is sought in the form 
\begin{equation}
\left( 
\begin{array}{c}
\Phi \left( r|r_{0}\right)  \\ 
\Psi \left( r|r_{0}\right) 
\end{array}%
\right) =\left( 
\begin{array}{c}
g\left( r|r_{0}\right) +C_{\Phi }\left( r_{0}\right) J_{\nu }\left(
k_{L}r\right)  \\ 
C_{\Psi }\left( r_{0}\right) J_{\nu }\left( k_{T}r\right) 
\end{array}%
\right) ,  \label{P8}
\end{equation}%
where $J_{\nu }\left( k_{L}r\right) $ and $J_{\nu }\left( k_{T}r\right) $
are the solutions of the homogeneous Bessel equation on the potentials.
Substituting (\ref{P8}) into the BC of (\ref{P7}) and appealing to (\ref{P4})%
$_{2}$ provides an algebraic system of equations on the coefficients $C_{\Phi
}\left( r_{0}\right) $ and $C_{\Psi }\left( r_{0}\right) .$ It has the form 
\begin{equation}
\mathbf{S}\left( R\right) \left( 
\begin{array}{c}
\Phi \left( R|r_{0}\right)  \\ 
\Psi \left( R|r_{0}\right) 
\end{array}%
\right) _{r_{0}<R}=\frac{2\mu }{R^{2}}\left[ \mathbf{\Lambda }\left( 
\begin{array}{c}
C_{\Phi }\left( r_{0}\right)  \\ 
C_{\Psi }\left( r_{0}\right) 
\end{array}%
\right) +bR^{2}\frac{\pi }{2}J_{\nu }\left( k_{L}r_{0}\right) \mathbf{\gamma 
}\right] =\mathbf{0}  \label{P10}
\end{equation}%
with 
\begin{equation}
\begin{array}{c}
\mathbf{\Lambda }=\left( 
\begin{array}{cc}
\left( \nu ^{2}-\frac{K_{T}^{2}}{2}-B_{L}\right) J_{\nu }\left( K_{L}\right) 
& j\nu \left( B_{T}-1\right) J_{\nu }\left( K_{T}\right)  \\ 
j\nu \left( B_{L}-1\right) J_{\nu }\left( K_{L}\right)  & -\left( \nu ^{2}-%
\frac{K_{T}^{2}}{2}-B_{T}\right) J_{\nu }\left( K_{T}\right) 
\end{array}%
\right) , \\ 
\ \mathbf{\gamma }=Y_{\nu }\left( K_{L}\right) \left( 
\begin{array}{c}
\nu ^{2}-\frac{K_{T}^{2}}{2}-B_{L}^{\left( Y\right) } \\ 
j\nu \left( B_{L}^{\left( Y\right) }-1\right) 
\end{array}%
\right) ,%
\end{array}
\label{P11}
\end{equation}%
where the use of (\ref{P11.1}) has been made, and $B_{L}^{\left( Y\right) }$
differs from $B_{L}$ by replacing $J_{\nu }$ with $Y_{\nu }$. Note that the
matrix (first-rank tensor) $\mathbf{\Lambda }$ results in (\ref{P10}) from
applying the boundary stress operator $\mathbf{S}\left( R\right) $ to the
solutions of the homogeneous equation. Hence $D=0,$ where%
\begin{equation}
D=\frac{\det \mathbf{\Lambda }}{J_{\nu }\left( K_{L}\right) J_{\nu }\left(
K_{T}\right) }=-\nu ^{4}+\nu ^{2}(B_{L}B_{T}+K_{T}^{2}+1)-\left( \frac{%
K_{T}^{2}}{2}+B_{L}\right) \left( \frac{K_{T}^{2}}{2}+B_{T}\right) ,
\label{P11.2}
\end{equation}%
is the dispersion equation for the eigenmodes of the traction-free cylinder.
By (\ref{P10}), 
\begin{equation}
\left( 
\begin{array}{c}
C_{\Phi }\left( r_{0}\right)  \\ 
C_{\Psi }\left( r_{0}\right) 
\end{array}%
\right) =\mathbf{-}bR^{2}\frac{\pi }{2}J_{\nu }\left( k_{L}r_{0}\right) 
\mathbf{\Lambda }^{-1}\mathbf{\gamma .}  \label{P10.1}
\end{equation}%
Inserting (\ref{P10.1}) along with (\ref{P9}) into (\ref{P8}) completes
finding the solution $\Phi \left( r|r_{0}\right) $ and $\Psi \left(
r|r_{0}\right) $ to (\ref{P7}).

\subsection{Displacement for a distributed source}

Once the solution $\Phi \left( r|r_{0}\right) $ and $\Psi \left(
r|r_{0}\right) $ to (\ref{P7}) has been found, it remains to take its
convolution with the radially distributed source $\sim $ $e^{-\alpha (R-r)}$
for obtaining the solution $\varphi \left( r\right) $ and $\psi \left(
r\right) $ to Eq. (\ref{P3}), and to invoke Helmholtz decomposition $\mathbf{%
u}=\mathbf{\nabla }\varphi +\mathbf{\nabla }\times \left( \psi \mathbf{e}%
_{z}\right) $ for finding $\mathbf{u}\left( r\right) .$ Because these two
are linear operations, they may be applied in the optional order. Thus 
\begin{equation}
\mathbf{u}\left( r\right) =\frac{1}{R^{2}}\int_{0}^{R}\left\{ \mathbf{\nabla 
}\Phi \left( r|r_{0}\right) +\mathbf{\nabla }\times \left[ \Psi \left(
r|r_{0}\right) \mathbf{e}_{z}\right] \right\} e^{-\alpha
(R-r_{0})}r_{0}dr_{0},  \label{P12}
\end{equation}%
where $\mathbf{G}\left( r|r_{0}\right) =\mathbf{\nabla }\Phi \left(
r|r_{0}\right) +\mathbf{\nabla }\times \left[ \Psi \left( r|r_{0}\right) 
\mathbf{e}_{z}\right] $ is the displacement response to a buried source $%
\mathbf{\nabla }\sigma ^{\left( T\right) }\left( r\right) \sim \mathbf{%
\nabla }\frac{\delta \left( r-r_{0}\right) }{r}.$ Note that application of
the boundary-stress differential operator to $g\left( r|r_{0}\right) $ (see (%
\ref{P8})) in the integrand of (\ref{P12}) taken with $r,r_{0}\rightarrow R$
provides an extra delta-function term, which compensates $\sigma ^{\left(
T\right) }\left( R\right) =-\beta T_{0}$ in the BC (\ref{P5}) for the
distributed source. 

In the context of laser ultrasonics, the important particular case is the
displacement at the surface $r=R.$ It can be written in the following closed
form:%
\begin{equation}
\mathbf{u}\left( R\right) =\left( 
\begin{array}{c}
u_{r} \\ 
u_{\theta }%
\end{array}%
\right) =IU_{0}\frac{K_{L}^{2}}{J_{\nu }\left( K_{L}\right) }\left( 
\begin{array}{c}
\nu ^{2}-B_{T}-\frac{K_{T}^{2}}{2} \\ 
j\nu \left(1- B_{T}\right)%
\end{array}%
\right) ,  \label{P18}
\end{equation}%
where 
\begin{equation}
I=\int_{0}^{1}J_{\nu }\left( K_{L}x\right) e^{-\alpha R(1-x)}x\mathrm{d}x\
\left( x=r/R\right) ,\ U_{0}=\frac{\beta T_{0}R}{2\mu D},  \label{P18.1}
\end{equation}%
and $D$ is given in (\ref{P11.2}).

Numerical simulations in \S 4.2 involve the model case of a buried source
located at $r_{0}=R-\epsilon $ tending to the surface. For this case, $%
\mathbf{u}\left( R\right) =\frac{1}{R^{2}}\mathbf{G}\left( R|R-\epsilon
\right) $ is given by (\ref{P18}) with $I=J_{\nu }\left( K_{L}\right) $ (the
equality is to within arbitrarily small $\epsilon /R$). Obviously, this case
may also be seen as a limit of vanishingly small optical penetration length $%
\alpha ^{-1}.$

\section{Numerical treatment and the simulation results}

\subsection{Inverse Fourier transform}

2D Fourier transform, discrete in circumferential number $\nu $ and
continuous in frequency $\omega ,$ has next been performed to obtain the
sought acoustic response in the time-space domain $\left\{ r,\theta
,t\right\} $. The calculation has been made for the radial component of
response at the surface, which is most usually the quantity measured by the
laser ultrasonics technique. Denote its above-derived value in the transform
domain as $u_{r}\left( R\right) =\widehat{u}_{r}(R,\nu ,\omega ),$ where $%
u_{r}\left( R\right) $ is given by Eq.(\ref{P18}). Fast Fourier Transform
algorithm in frequency has been applied, with a small imaginary part added
to $\omega $ ($\omega ^{\ast }=\omega -\jmath \delta $ where $\delta >0,$
e.g. \cite{Weaver:1996}) in order to avoid zeros of the dispersion equation $%
D=0,$ see (\ref{P18.1}) and the remark to (\ref{P11.2}). On taking Fourier
transform in $\nu ,$ the use has been made of the fact that, by (\ref{P18}), 
$\widehat{u}_{r}(r,\nu ,\omega )$ is an even function of $\nu .$ As a result
the sought response $u_{r}(R,\theta ,t)$ is described by the formula%
\begin{equation}
u_{r}(R,\theta ,t)=\frac{e^{-\delta t}}{\pi }\int_{-\infty }^{\infty
}\left\{ \sum_{\nu =0}^{\infty }\varepsilon _{\nu }\widehat{u}_{r}(R,\nu
,\omega )cos(\nu \theta )\right\} e^{-\jmath \omega t}d\omega ,
\label{eq:inverse_scheme2}
\end{equation}%
where $\varepsilon _{\nu }$ is the Neumann factor equal to $1$ if $\nu =0$
and to $2$ otherwise.

The series in (\ref{eq:inverse_scheme2}) is non-absolutely convergent \cite%
{Gespa:1987}. Its convergence in $\nu $ depends on $\omega $ and hence so
does the truncation order $\nu _{m},$ which is supposed to satisfy the
criterion%
\begin{equation}
\frac{|\widehat{u}_{r}(R,\nu _{m},\omega )|}{\sum_{\nu =0}^{\nu _{m}}|%
\widehat{u}_{r}(R,\nu ,\omega )|}<\varepsilon ,  \label{eq:criterion}
\end{equation}%
where $\varepsilon $ is the error bound. It is, however, known that usually
the sum in question is mainly due to the first terms of the sum. For the
frequency band used in our case, it was verified that taking a fixed
truncation order $\nu _{m}$ such that keeps about 200 terms of the series
ensures that the committed error does not exceed $10^{-5}$.

\subsection{Results of numerical simulations}

Numerical simulations have been implemented for the opaque aluminium rod and
the optically absorptive NG5 coloured glass rod, both of the diameter
5 mm. The values of their elastic and thermal constants are listed in Table~\ref%
{tab:table1} (the notations are explained below (\ref%
{eq::coupled_thermoelastic_equations})). Assuming the laser wavelength 355
nm, the absorption length for NG5 coloured glass is $\alpha ^{-1}=$ 0.8 mm.
The typical values of the laser beam-width $b=0.15$ mm and of the pulse
duration 5 ns are taken. The radial component of the surface displacement $%
u_{r}(R,\theta ,t)$ is supposed to be detected at the point $\left( r,\theta
\right) =\left( R,\pi \right) $ which is diametrically opposite to the
generation point $\left( R,0\right) $. Fourier transform in $\omega $ has
been performed with the numerical parameter $\delta =0.035,$ see (\ref%
{eq:inverse_scheme2}).

\begin{table}[h]
\caption{\label{tab:table1}Physical properties of Al and NG5 glass used in simulations.}
\begin{center}
\lineup
\begin{tabular}{*{6}{c}}
\br                              
 &$\rho$ [g cm$^{-3}$]&$\lambda$ [GPa]&\m $\mu$ [GPa]&\m $\alpha_T$
 [K$^{-1}$]&$c_p$ [J kg$^{-1}$ K$^{-1}$]\cr
\mr
\0\0Al &2.69 &56   &26.5 &2.5e-6 &902 \cr
\0\0NG5&2.31 &17.4 &22.9 &6.5e-6 &700\cr
\br
\end{tabular}
\end{center}
\end{table}

\begin{figure}[htbp!]
\begin{center}
\includegraphics[width=160mm,angle=0]{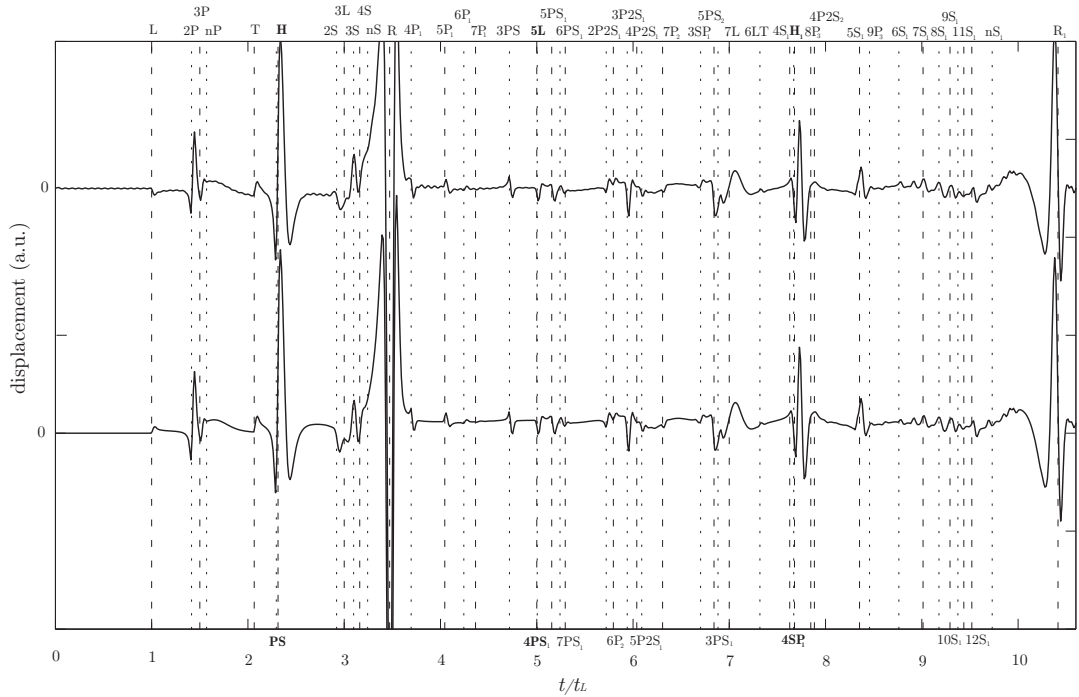}
\end{center}
\caption[Simulations of a buried source and a dipole
force located at the surface]{Acoustic response of the aluminium rod
  computed for the model of a buried source tending to the surface
  (upper curve) and for Scruby's model of a surface source (lower
  curve). The notations for wave arrivals are explained in text.}
\label{fig:P_T}
\end{figure}

\begin{figure}[htbp!]
\begin{center}
\includegraphics[width=160mm]{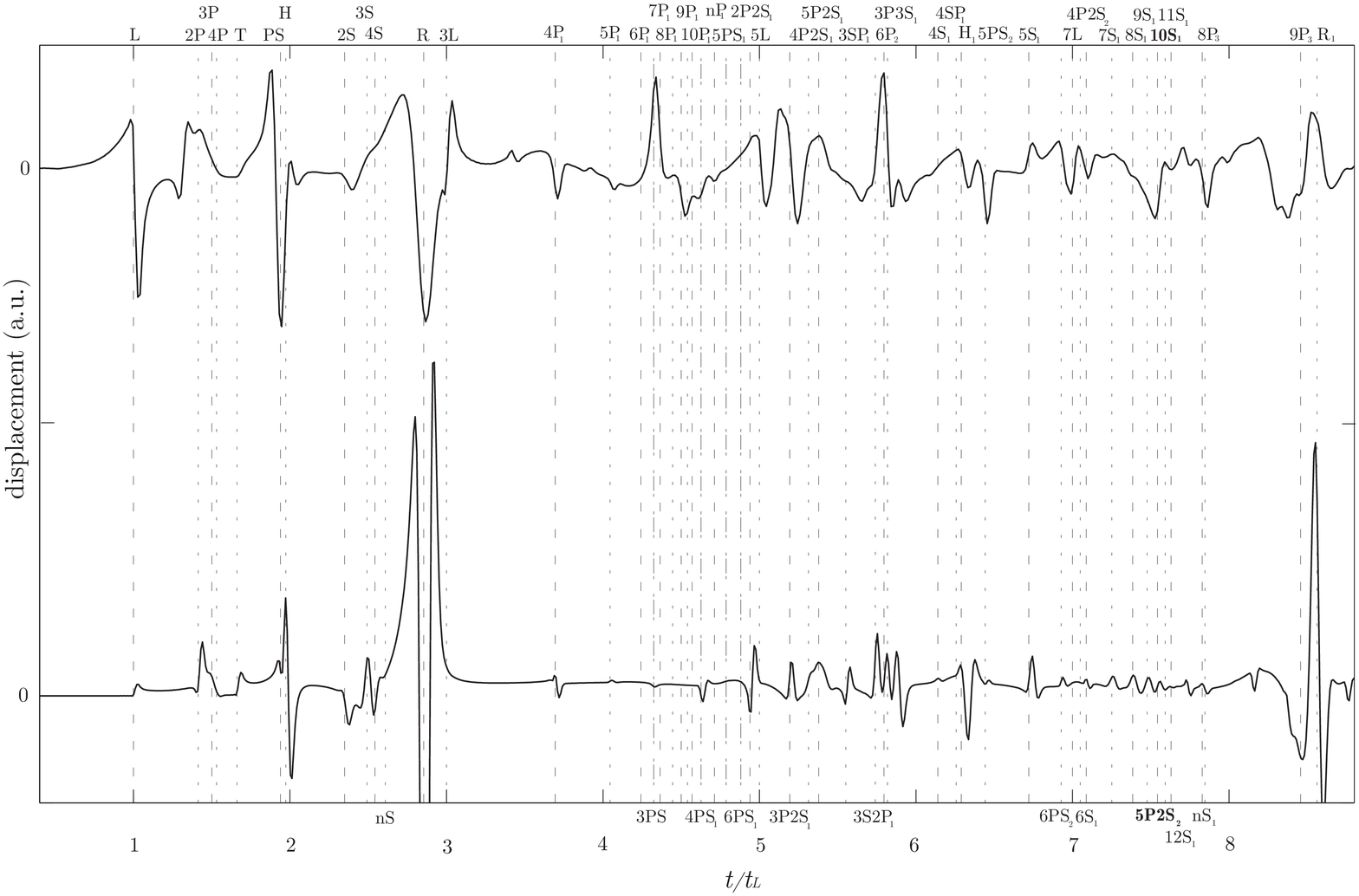}
\end{center}
\caption[Experimental signals compared to numerical
simulations]{Acoustic response of the optically absorptive NG5 coloured glass
  rod. The upper curve is calculated for the radially
  distributed source taking into account optical penetration. The
  lower curve is calculated from Scruby's model of a surface source
  for opaque materials.}
\label{fig:NG5_numerical}
\end{figure}

The results of numerical simulations are presented in Figs.\ref{fig:P_T} and %
\ref{fig:NG5_numerical}. The wave arrivals are marked by the dashed lines.
They are identified and denoted according to the nomenclature proposed in 
\cite{Pan:2004} on the basis of a ray trajectory analysis. The notation $nL$
and $nT$ corresponds to the arrivals of longitudinal and transverse bulk
modes, travelling along the straight diameter path and undergoing $n$ normal
reflections from the cylinder surface at the edge points. The notation $nP$
and $nS$ also corresponds to, respectively, the longitudinal and transverse
bulk modes, but these ones arrive at the detection point after a broken path
resulting from $n$ oblique reflections without modal conversion. The wave
arrivals, involving modal conversion at oblique reflections, are denoted by $%
mPnS_{r},$ where the subscript $r$ is the number of round trips about the
center made within the given path. The Rayleigh-wave arrival is denoted by $%
R_{r}$. The notation $H_{r}$ is used for the wave arrival due to the head
wave, i.e., due to the transverse mode radiated by the skimming longitudinal
wave in the direction of the critical angle $\varphi _{c}=\arcsin \sqrt{2\mu
/\left( \lambda +2\mu \right) }.$

Fig.\ref{fig:P_T} compares two simulations of the laser-generated acoustic
response $u_{r}(R,\pi ,t)$ for the opaque aluminium cylinder.
 The upper
curve is computed by means of the developed procedure reduced to the model
case such that assumes a radially concentrated buried source tending to the
cylinder surface or, equally, a vanishingly small optical penetration length
(see the closing remark in \S\ 3.3). The lower curve is computed by means of
Scruby's model of the surface dipole source \cite{Scruby:1990}. Precise
agreement of the arrival times and waveform shapes of both signals is
observed.

Fig.\ref{fig:NG5_numerical} presents the numerical results for the
optically absorptive NG5 coloured glass cylinder. The upper curve shows the acoustic
response $u_{r}(R,\pi ,t)$ computed from Eqs. (\ref{P18})-(\ref%
{eq:inverse_scheme2}) for the radially distributed source, which takes into
account marked enough optical penetration of the given material ($\alpha
R\approx 3$). This curve is again compared with the 'would-be' signal
calculated for this case on the basis of Scruby's surface-source model
(lower curve). It is seen that while the arrival times for both signals are
obviously in good accordance, there is an evident discrepancy in the shape
of the waveforms. For instance, the first longitudinal-wave arrival $L$ on
the upper curve has a characteristic bipolar broad shape, which is not
observed on the lower curve. Another striking dissimilarity is that the
bulk-wave and surface-wave arrivals on the upper curve have more or less
similar amplitudes, which is in contrast to the predominant peaks of the
first Rayleigh-wave arrivals $R$ and $R_{1}$ on the lower curve. Altogether,
the difference between these two curves highlights the effect of optical
penetration, which Scruby's surface-source model is not intended to
describe.

\section{Conclusions}

Acoustic response of an isotropic cylinder to a laser line pulse has been
simulated in the framework of 2D semi-analytical model, which takes into
account optical penetration into the bulk of optically absorptive cylinder material.
The waveforms calculated for the aluminium rod with a vanishingly small
penetration coincide, as expected, with those obtained from Scruby's model
of a surface dipole source. At the same time, calculation for the glass rod
with large enough penetration length has demonstrated an essential
difference of the waveform shapes relatively to the signal calculated for
the same material by way of the surface-source model.

The semi-analytical procedure developed in this paper paves the way to
quantitative analysis of experimental results of laser ultrasonics
generation in cylindrical samples made of optically absorptive materials. First
experimental data have recently been obtained and will be reported elsewhere.

\section*{References}


\providecommand{\newblock}{}

\end{document}